# Scalable Gate-Defined Majorana Fermions in 2D p-Wave Superconductors

*Ji Ung Lee\**

College of Nanoscale Science and Engineering, SUNY-Polytechnic Institute, Albany, NY 12203, United States

ABSTRACT

We provide a conceptual framework for developing a scalable topological quantum computer.  It relies on forming Majorana fermions using circular electronic gates in two-dimensional p-wave superconductors. The gates allow the precise control of the number, position, and dynamics of Majorana fermions.  Using an array of such gates, one can implement the full features of topological quantum computation, including the braiding and fusion of Majoranas in space-time. The gates serve two purposes: They modulate the chemical potential locally to turn a topological superconductor into a normal conductor, and they are used to move the Majoranas in space-time. With a perpendicular magnetic field, the normal region localizes a quantum of magnetic flux. Under these conditions, the boundary between the normal region and the superconducting region supports a single zero-energy Majorana bound state.  The localized zero mode is sufficiently separate from other states and can be dragged by sequentially applying voltages to the adjacent gates to implement quantum computation. We briefly describe the fabrication process to construct the device and determine key properties from experimentally determined parameters. The digital qualities of topological protection provide intrinsic immunity to the inevitable fabrication nonuniformities.

INTRODUCTION

Recent advances in topological quantum computation have focused on finding signatures of Majorana fermions localized in one-dimensional (1D) nanowires[1-4]. A Majorana fermion possesses the property that it is its own anti-particle. They possess non-Abelian exchange statistics, which can be used for quantum information processing that is immune to decoherence[5-7]. One of the key demonstrations of Majorana-based topological quantum computation is the process of braiding and fusion of Majorana fermions in space-time, which is easier to accomplish in 2D. While device concepts have been put forth to exchange Majoranas in 1D[8], only a handful such concepts have been discussed for 2D systems in a manner that can scale[9].

Here, we propose the elements of a topological quantum computer that uses purely electrical means to localize and drag Majorana fermions in 2D. Specifically, we use nanoscale electronic gates instead of using mechanical means[10, 11] to manipulate the Majorana fermion localized in a quantum of magnetic flux. The approach relies on recent advances in topological superconducting materials [12-16] and planar fabrication processes. Most importantly, the proposed approach can scale and is able to explore a large number of entangled Majorana fermions.

DISCUSSION

We begin by summarizing the analysis given elsewhere[6, 7, 17, 18] to help guide the construction of the device given below. The starting point for our discussion is a lattice model for a spinless two-dimensional p-wave superconductor. For simplicity, we consider a square lattice with a lattice constant $a$ of a hypothetical p-wave superconductor, which is needed to realize a topological superconductor. Although most superconductors possess an s-wave order parameter, certain superconductors possess an effective p-wave symmetry at surfaces [12-16]. To preserve the anticommutative property of fermions in a p-wave superconductor, the order parameter must have odd parity. We consider a px+ipy symmetry for the order parameter.

The real-space Hamiltonian for the square lattice is given as

$$H = \sum_{mn}[-t(c^+_{m+1,n}c_{mn} + c^+_{m,n+1}c_{m,n}) + h.c. - (\mu - 4t)c^+_{m,n}c_{m,n} + \Delta c^+_{m+1,n}c^+_{m,n} + i\Delta c^+_{m,n+1}c^+_{m,n} + h.c.] \tag{1}$$

where $c^+_{m,n}$ and $c_{m,n}$ creates and annihilates, respectively, a particle on the lattice site (*m*,*n*); *t* is the nearest neighbor hopping energy, μ is the chemical potential, and $\Delta$ is the superconducting pairing gap. The chemical potential is offset by $4t$ in anticipation of forming a topological phase when μ > 0. We cast $H$ in the Bogoliubov-de Gennes (BdG) form. This doubles the size of the matrix and introduces particle-hole symmetry. The symmetry can lead to a doubly degenerate state at zero-energy, where we look for Majorana fermions. We take the usual approach by moving to the Fourier space assuming translational invariance. The resulting $H_{BdG}$ is given as

$$H_{BdG} = \sum_k \frac{1}{2}(c^+_k \; c_{-k})H(k)\begin{pmatrix}c_k \\ c^+_{-k}\end{pmatrix}$$

which can be diagonalized by a unitary transformation known as the Bogoliubov-Valatin transformation. For $k \cong 0$

$$H(k) \approx \begin{bmatrix} -\mu & 2i\Delta a(k_x + ik_y) \\ -2i\Delta a(k_x - ik_y) & \mu \end{bmatrix} \quad (2).$$

The single-particle spectrum of $H(k)$ is fully gapped because there is no boundary. $H(k)$ can be cast in terms of the Dirac equation $H(k) = \boldsymbol{h}(k) \cdot \boldsymbol{\sigma}$ where $\boldsymbol{\sigma}$ is a vector formed of Pauli spin matrices. This is the form that is often used to determine the topological phases and Chern numbers. For example, the diagonal terms are the components of $\sigma_x$ and are equivalent to the mass term in the Dirac equation, with the unusual property that it ($-\mu$) becomes negative when $\mu > 0$. Thus $\mu > 0$ represents the topological superconducting (TSC) phase. Therefore, a boundary between a normal region or a region with $\mu < 0$ and a topological superconductor ($\mu > 0$) can host a chiral edge state. Assuming this boundary is located at $x = 0$ ($\mu > 0$ for $x > 0$), with translational symmetry along the y-axis, we can solve for the edge state by substituting $k_x = \frac{1}{i}\frac{d}{dx}$ and solving for the eigenstate $\psi(x,y)$. The solution for $E = 0$ follows the Jackiw-Rabbi form. For $x > 0$ we have

$$\psi(x,y) = \psi(x)e^{ik_y y}; \quad \psi(x) = \begin{pmatrix} 1 \\ -1 \end{pmatrix} \exp\left(\frac{-1}{2\Delta a}\int_0^x dx' \mu(x')\right).$$

If $\mu$ is spatially uniform, $2\Delta a/\mu$ is the spatial extent of the zero mode in the x-axis. It depends inversely on the chemical potential $\mu$, which can be tuned by a gate to control the interaction between Majorana fermions, as we discuss below. The zero mode is chiral with a massless linear dispersion relation given by $E_y = \Delta a k_y$ for small $k$, with velocity $v = \Delta a/\hbar$. It is similar to that of graphene except $\Delta$ is replaced by $t$ in this case. Since $\Delta$ is much smaller than the hopping parameter $t$ for graphene, the velocity of the chiral state is much less than the Fermi velocity of graphene.

The linear dispersion in $k_y$, together with the spectrum from equation Eq. (2), gives the band structure shown in Fig. 1(a) for a topological superconductor in proximity with a trivial material.

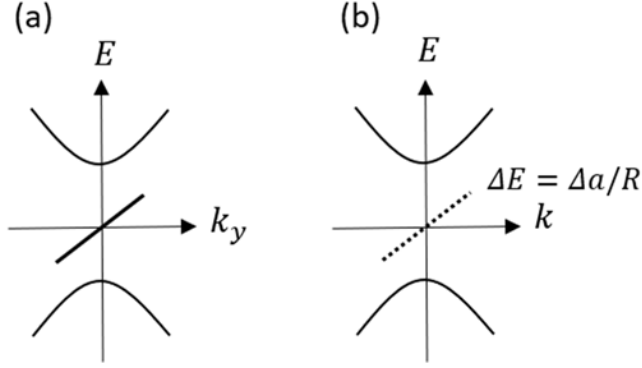

Figure 1: (a) A single chiral linear dispersion within the superconductig gap at a boundary between a trivial region and a topological superconductor. (b) The effect of introducing a circular boundary of radius R supporting a quantum of magnetic flux. The chiral dispersion relation now becomes discrete and hosts a Majorana zero-energy bound state that is seprated by energy $\Delta E$ from the other states.

The zero-energy solution at $k_y=0$ in Fig. 1(a) hosts a Majorana zero mode $\gamma$. To see this, we take the Bogoliubov-Valatin transformation and define

$$\gamma = e^{i\pi/2} \int dr^2 e^{ik_y y}(\exp(\tfrac{-1}{2\Delta a}\int_0^x dx'\mu(x'))[e^{-i\pi/2}\psi(r) + e^{i\pi/2}\psi(r)^+]$$

where $\psi(r)$ and $\psi(r)^+$ are field operators. The overall phase $e^{i\pi/2}$ is not important and we can ignore it. Then, $\gamma$ is the field operator that satisfies the Majorana condition $\gamma = \gamma^+$ provided $k_y = 0$ and $E = 0$.

Unfortunately, the Majorana zero mode is within the continuum of the linear dispersion relation, and its lifetime is expected to be very short. In 1D, the problem is solved since the y-axis does not enter into the problem, and one obtains a pair of localized Majorana zero modes in the nanowire. In 2D, one may be tempted to consider a periodic boundary condition in the y-axis, as would apply for a circular boundary. Moving to polar coordinates, however, one can show that because of the orbital angular momentum of the chiral state, a circular boundary does not result in a zero-energy mode. A remedy to this problem is found by inserting a quantum of magnetic flux through a trivial region of radius $R$, as illustrated in Fig. 2. This situation is analogous to an Abrikosov vortex in a p-wave superconductor, which does host a zero-energy Majorana mode at its core. In the Abrikosov vortex, the zero-point energy of the core offsets the finite energy due to the angular momentum, resulting in discrete states $E_n = n\Delta a/R$ where $n$ is an integer. These discrete levels are the Caroli–de Gennes–Matricon bound states of a vortex[19] for a p-wave superconductor. The vortex, therefore, hosts a Majorana mode with exactly zero-energy.

For the vortex $R \sim \xi$, the coherence length of the superconductor[7]. These discrete states are shown in Fig. 1(b). Because of the particle-hole symmetry, the solution at $E = 0$ is insensitive to the actual profile of $\mu(x)$. Controlling the density, location, and motion of Abrikosov vortices, however, will be problematic. We show below that $\Delta E = \Delta a/R$ can be large in a gate-defined normal core so that the exponential protection is large, with none of the constraints of Abrikosov vortices.

Instead of using Abrikosov vortices, our approach is to create a gate-defined normal core, as shown in Fig. 2. The approach is different than those proposed by others using vortices[20, 21]. Here, we rely on recent demonstrations showing that superconductivity can be tuned to an insulating state by an electronic gate[22]. This feature should be a general property for 2D materials. Since a void or a normal region embedded in a superconductor can also admit a quantum of magnetic flux, the analysis is identical to the one already developed for Abrikosov vortices discussed above except for the location of the zero mode. The location of the Majorana zero mode is defined either by the gate boundary or by the coherence length $\xi$ of the superconductor, whichever is greater. We assume for our discussion that $R > \xi$. The size of the vortex is defined by the London penetration depth λ, which we assume is much larger than $R$.

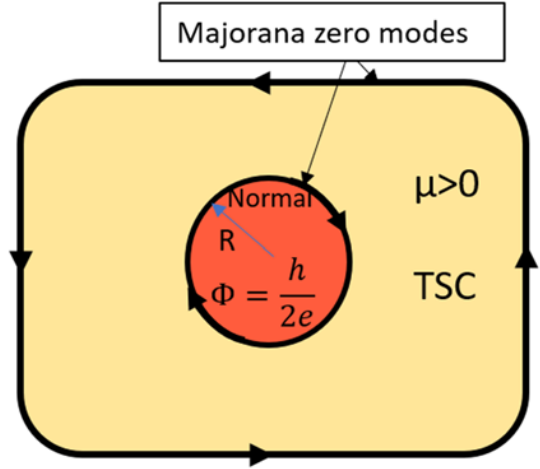

*Figure 2 : A topological superconductor (TSC) containing a normal core of radius R with a quantum of magnetic flux threading the region. Here, the normal region is created by a gate, as shown in Fig. 3. The boundary of the normal core hosts a single chiral Majorana zero mode. Its partner is located at the outside edge of the superconductor and has the opposite chirality, as denoted by the arrows.*

The proposed system can support either a type I or type II superconductor. For the former, the applied magnetic field must be less than the critical field of the superconductor. For the latter, the magnetic field must be below the lower critical field to prevent the formation of unwanted Abrikosov vortices. Due to the Meissner effect, flux will be expelled in the bulk of the superconducting film and concentrate in the normal regions defined by the gates. Since Majorana fermions must come as a pair, there is another Majorana zero mode with the opposite chirality. It is spatially separate from the one at the edge of the normal core and is located at the outside edge of the superconductor, as shown in Fig. 2. The outside edge states are present whenever there is an odd number of zero modes within the film. Whenever there is an even number of Majorana zero modes in the film, the outside edge mode no longer hosts a zero-energy state, and the Majorana pairs reside within the bulk of the film.

With a dense array of circular gates, one can implement the full features of Majorana-based quantum computers. The proposed structure is shown in Fig. 3(a). A 2D p-wave superconductor is gated from above and below. A top global gate tunes µ to the superconducting topological phase and controls the spatial extent of the zero modes to minimize the hybridization between neighboring Majoranas. One also needs to apply a perpendicular magnetic field (not shown) to create the Majoranas. A voltage applied to a circular gate creates a normal region that traps a magnetic flux quantum. Sequentially applying voltages to the neighboring gates while turning off the voltage to the one creating the zero mode moves the Majorana in space-time.

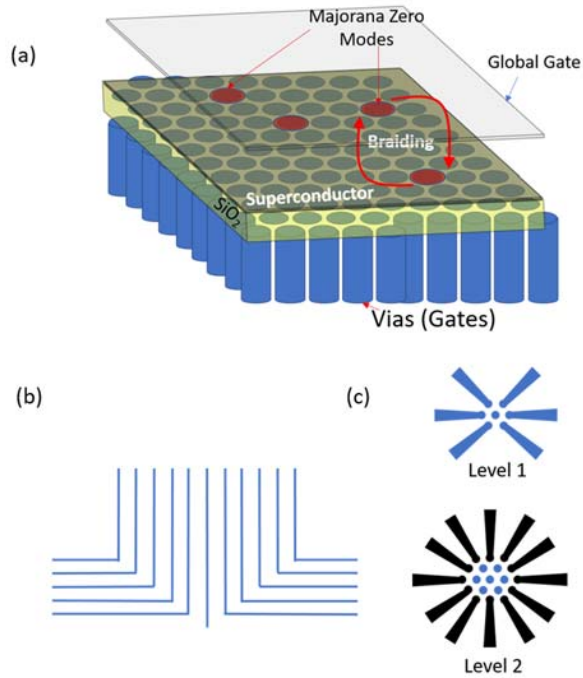

*Figure 3: (a) An array of individually addressable circular electrical gates(vias) to define and manipulate Majorana fermions in a 2D p-wave superconductor. The top global gate modulates the chemical potential so that the superconductor is in the topological phase. The bottom circular gates are biased to turn regions into normal conductors (shown in red). In the presence of a vertical magnetic field (not shown), each normal regions traps a quantum of magnetic flux, which hosts a Majorana zero mode. (b) Cross-sectional image of the vias: To fabricate the individually addressable vias, a multilevel fabrication process can be used to form the vias and connect them with radially-directed metal electrodes. (c) The top-down image of the vias and electrodes for the bottom two layers.*

The close-packed structure is needed so that there is a seamless coupling of the Majorana mode formed between two neighboring gates. Fortunately, electrostatics favor such bias conditions. We illustrate in Fig. 4 (a)-(c) the transfer of a Majorana zero mode between two neighboring gates. For example, when two neighboring gates are biased with the same voltage, the potential profile at the superconducting plane between two neighboring gates becomes smooth. This is illustrated in Fig. 4(b). Since the size of the vortex extends by $\lambda \gg R$, one can readily see that the intermediate bias condiction shown in Fig. 4(b) represents the same vortex. Furthermore, the topological nature of the Majorana zero mode, which is pinned at zero-energy, ensures that it can be moved from one gated region to the next. The topological protection also ensures that the process is immune to the inevitable variation in the gate geometry, as illustrated in Figs. 4(d)-(f).

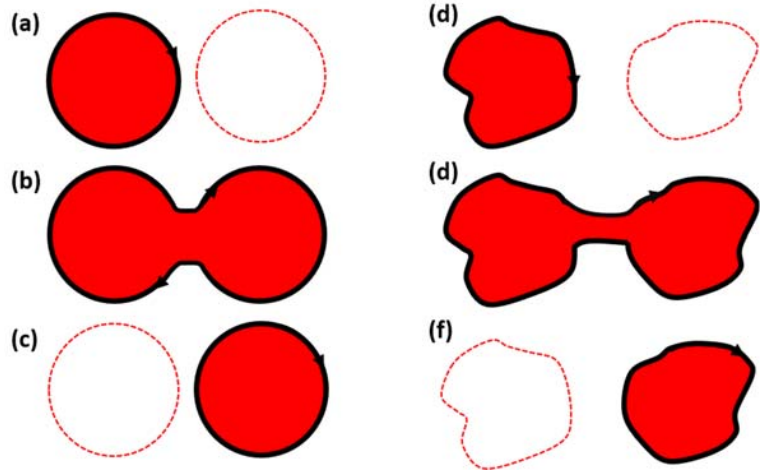

*Figure 4: Moving a single normal core. (a)-(c) depict a single normal core containing a Majorana zero mode moving from the left gate to the right gate. In the intermediate region (b) when two gates are turned on, the zero mode extends over two neighboring gates. Since the size of the vortex extends by $\lambda \gg R$, (a)-(c) depict the same vortex. (d)-(f) represent the same process as in (a)-(c) but is depicted using non-uniform gate geometries. Due to topological protection, the zero mode is immune to the inevitable process variations that yield non-uniform gate geometries as shown in (d)-(f).*

The biggest challenge is in fabricating the array of individually addressable, dense circular gates with nanometer-scale diameter. One way is to use anodized aluminum oxide nanopores as a template. Pore sizes as small as $R \sim 2.5$ nm in radius have been achieved[23, 24]. The nanopores can be filled with a wide range of metals using an electrochemical plating process[25] to create vertical metallic nanowires. To connect to the individual nanowires, we can follow the procedure we discuss next based on an alternative technique to create the vertical electrodes.

An alternative technique is to use the metallization process developed for the microelectronics industry. There, vias and Damascene processes are used to connect transistors to different metal layers. The via electrodes can be built-up sequentially using a multilevel fabrication process and connected using outward-directed metal leads. In Fig 3(b), we show a cross-sectional image of the vias and metal leads fabricated at different levels. From the center via, a growing number of radially positioned vias can be fabricated at each level (Fig. 3c). At each level, planar metal leads that extend radially outward connect to the individual via electrodes. By building the metal leads at different levels, they are allowed to cross, allowing a high density of vias.

Modern fabs can pattern vias that are less than 20nm in diameter using EUV lithography[26], while electron beam lithography can pattern smaller features. The topological protection of the zero mode guarantees that it will be insensitive to the inevitable variations in the via diameter and shape. We show the robustness of the Majorana zero mode by comparing to experimentally determined Caroli–de Gennes–Matricon levels for Abrikosov vortices in a Fe-based superconductor, for which $E_n \sim n(0.65 meV)$ with $\xi \sim 2.5 nm$ were measured[16]. Scaling to $R = 10 nm$, we would achieve $\Delta E = E_1 - E_0 = 0.16 meV$ between the Majorana zero mode and the $n=+/-1$ angular momentum states. At a refrigeration temperature of $T = 10 mK$, the thermal barrier is $\exp(\Delta E/K_B T) > 10^{80}$ where $K_B T$ is thermal energy. As a reference, this value is more than 70 orders larger than the comparable values being explored in Josephson junction-based transmon qubits[27] at the same temperature. A large energy difference is still expected even as the Majorana zero mode is extended between two neighboring gates while it is being moved.

CONCLUSION

In summary, we provide a framework for developing Majorana-based topological quantum computers in 2D p-wave superconductors. The Majorana fermions are defined by electrical gates, which are arrayed and addressed individually to allow braiding and fusion of Majoranas, the central feature of topological computation. The gates create topologically trivial regions embedded in a topological superconductor. The trivial regions can host Majorana fermions when they trap a quantum of magnetic flux. Using experimentally realizable geometries, we show that the Majorana zero mode possesses large exponential protection from other states to make the proposed approach technologically viable. Furthermore, the topological nature that pins the bound state to zero-energy provides intrinsic digital immunity to fabrication nonuniformities. Finally, the proposed approach is compatible with CMOS processing in anticipation of the need to integrate classical and quantum computers.

**AUTHOR INFORMATION**
**Corresponding Author**
*jlee1@sunypoly.edu

**REFERENCES**

[1]   V. Mourik, K. Zuo, S. M. Frolov, S. R. Plissard, E. P. A. M. Bakkers, and L. P. Kouwenhoven, "Signatures of Majorana Fermions in Hybrid Superconductor-Semiconductor Nanowire Devices," *Science,* vol. 336, no. 6084, pp. 1003-1007, 2012, doi: 10.1126/science.1222360.
[2]   S. M. Albrecht *et al.*, "Exponential protection of zero modes in Majorana islands," *Nature,* vol. 531, no. 7593, pp. 206-209, 2016/03/01 2016, doi: 10.1038/nature17162.
[3]   Ö. Gül *et al.*, "Ballistic Majorana nanowire devices," *Nat Nanotechnol,* vol. 13, no. 3, pp. 192-197, 2018/03/01 2018, doi: 10.1038/s41565-017-0032-8.
[4]   H. Zhang *et al.*, "Quantized Majorana conductance," *Nature,* vol. 556, no. 7699, pp. 74-79, 2018/04/01 2018, doi: 10.1038/nature26142.

[5] A. Y. Kitaev, "Fault-tolerant quantum computation by anyons," *Annals of Physics,* vol. 303, no. 1, pp. 2-30, 2003/01/01/ 2003, doi: https://doi.org/10.1016/S0003-4916(02)00018-0.

[6] C. Nayak, S. H. Simon, A. Stern, M. Freedman, and S. Das Sarma, "Non-Abelian anyons and topological quantum computation," *Reviews of Modern Physics,* vol. 80, no. 3, pp. 1083-1159, 09/12/ 2008, doi: 10.1103/RevModPhys.80.1083.

[7] J. Alicea, "New directions in the pursuit of Majorana fermions in solid state systems," *Reports on Progress in Physics,* vol. 75, no. 7, p. 076501, 2012/06/28 2012, doi: 10.1088/0034-4885/75/7/076501.

[8] D. Aasen *et al.*, "Milestones Toward Majorana-Based Quantum Computing," *Physical Review X,* vol. 6, no. 3, p. 031016, 08/03/ 2016, doi: 10.1103/PhysRevX.6.031016.

[9] C. W. J. Beenakker, A. Brabsch, and Y. Herasymenko, "Electrical detection of the Majorana fusion rule for chiral edge vortices in a topological superconductor," *SciPost Phys.,* vol. 6, p. 022, 2019.

[10] A. Kremen, S. Wissberg, N. Haham, E. Persky, Y. Frenkel, and B. Kalisky, "Mechanical Control of Individual Superconducting Vortices," *Nano Lett,* vol. 16, no. 3, pp. 1626-1630, 2016/03/09 2016, doi: 10.1021/acs.nanolett.5b04444.

[11] J.-Y. Ge *et al.*, "Nanoscale assembly of superconducting vortices with scanning tunnelling microscope tip," *Nature Communications,* vol. 7, no. 1, p. 13880, 2016/12/09 2016, doi: 10.1038/ncomms13880.

[12] S. Charpentier *et al.*, "Induced unconventional superconductivity on the surface states of $Bi_2Te_3$ topological insulator," *Nature Communications,* vol. 8, no. 1, p. 2019, 2017/12/08 2017, doi: 10.1038/s41467-017-02069-z.

[13] D. Wang *et al.*, "Evidence for Majorana bound states in an iron-based superconductor," *Science,* vol. 362, no. 6412, pp. 333-335, 2018, doi: 10.1126/science.aao1797.

[14] J.-P. Xu *et al.*, "Experimental Detection of a Majorana Mode in the core of a Magnetic Vortex inside a Topological Insulator-Superconductor $Bi_2Te_3$/$NbSe_2$ Heterostructure," *Phys Rev Lett,* vol. 114, no. 1, p. 017001, 01/07/ 2015, doi: 10.1103/PhysRevLett.114.017001.

[15] S. Zhu *et al.*, "Nearly quantized conductance plateau of vortex zero mode in an iron-based superconductor," *Science,* vol. 367, no. 6474, pp. 189-192, 2020, doi: 10.1126/science.aax0274.

[16] L. Kong *et al.*, "Half-integer level shift of vortex bound states in an iron-based superconductor," *Nature Physics,* vol. 15, no. 11, pp. 1181-1187, 2019/11/01 2019, doi: 10.1038/s41567-019-0630-5.

[17] N. Read and D. Green, "Paired states of fermions in two dimensions with breaking of parity and time-reversal symmetries and the fractional quantum Hall effect," *Physical Review B,* vol. 61, no. 15, pp. 10267-10297, 04/15/ 2000, doi: 10.1103/PhysRevB.61.10267.

[18] B. A. Bernevig and T. L. Hughes, *Topological Insulators and Topological Superconductors*. Princeton, NJ: Princeton University Press, 2013.

[19] C. Caroli, P. G. De Gennes, and J. Matricon, "Bound Fermion states on a vortex line in a type II superconductor," *Physics Letters,* vol. 9, no. 4, pp. 307-309, 1964/05/01/ 1964, doi: https://doi.org/10.1016/0031-9163(64)90375-0.

[20] C. W. J. Beenakker, P. Baireuther, Y. Herasymenko, I. Adagideli, L. Wang, and A. R. Akhmerov, "Deterministic Creation and Braiding of Chiral Edge Vortices," *Phys Rev Lett,* vol. 122, no. 14, p. 146803, 04/11/ 2019, doi: 10.1103/PhysRevLett.122.146803.

[21] X. Ma, C. J. O. Reichhardt, and C. Reichhardt, "Braiding Majorana fermions and creating quantum logic gates with vortices on a periodic pinning structure," *Physical Review B,* vol. 101, no. 2, p. 024514, 01/21/ 2020, doi: 10.1103/PhysRevB.101.024514.

[22] V. Fatemi *et al.*, "Electrically tunable low-density superconductivity in a monolayer topological insulator," *Science,* p. eaar4642, 2018, doi: 10.1126/science.aar4642.

[23] H. Lira and R. Paterson, "New and modified anodic alumina membranes: Part III. Preparation and characterisation by gas diffusion of 5 nm pore size anodic alumina membranes," *Journal of Membrane Science,* vol. 206, pp. 375-387, 08/31 2002.

[24] M. J. Pellin *et al.*, "Mesoporous catalytic membranes: Synthetic control of pore size and wall composition," *Catalysis Letters,* vol. 102, no. 3, pp. 127-130, 2005/08/01 2005, doi: 10.1007/s10562-005-5843-9.

[25] A. Ghaddar, J. Gieraltowski, and F. Gloaguen, "Effects of anodization and electrodeposition conditions on the growth of copper and cobalt nanostructures in aluminum oxide films," *Journal of Applied Electrochemistry,* vol. 39, no. 5, pp. 719-725, 2009/05/01 2009, doi: 10.1007/s10800-008-9715-z.

[26] S. Park *et al.*, "EUV contact holes and pillars patterning," *Proceedings of SPIE - The International Society for Optical Engineering,* vol. 9422, 03/16 2015, doi: 10.1117/12.2085920.

[27] J. M. Gambetta, J. M. Chow, and M. Steffen, "Building logical qubits in a superconducting quantum computing system," *npj Quantum Information,* vol. 3, no. 1, p. 2, 2017/01/13 2017, doi: 10.1038/s41534-016-0004-0.